\documentclass[prb,twocolumn, showpacs, superscriptaddress]{revtex4}
\usepackage{amssymb}
\usepackage{amsmath}
\usepackage{bbm}
\usepackage{graphicx}

\begin{document}

\author{Philipp Werner}
\affiliation{Theoretische Physik, ETH Zurich, 8093 Zurich, Switzerland}
\author{Emanuel Gull}
\affiliation{Theoretische Physik, ETH Zurich, 8093 Zurich, Switzerland}
\author{Andrew J. Millis}
\affiliation{Columbia University, 538 West, 120th Street, New York, NY 10027, USA}
\title{Metal-Insulator phase diagram and orbital selectivity in  3-orbital models with rotationally invariant Hund coupling}
\date{\today}

\hyphenation{}

\begin{abstract}
A three band model containing the essential physics of  transition metal oxides with partially filled $t_{2g}$ shells is solved in the single-site dynamical mean field approximation, using the full rotationally invariant Slater-Kanamori interactions. We compute the metal-Mott insulator phase diagram in the space of chemical potential and interaction strength, determine the response of the different phases to perturbations which break the orbital symmetry, and establish the regimes in which an orbital selective Mott phase occurs. The results are compared to data on titanates, ruthenates, vanadates and C$_{60}$.
\end{abstract}


\maketitle

\section{Introduction}

The physics of  strong (electronic) correlations plays a central role in modern-day condensed matter physics.\cite{Imada98,Science00} The essence of this problem is the competition between the repulsive interactions felt by electrons in transition metal $d$-orbitals or lanthanide/actinide $f$ orbitals and the itineracy arising from hybridization with other orbitals in the material. For an atom in free space, the $d$ and $f$ shells have respectively a $5$ and $7$-fold orbital degeneracy and when the orbitals are partially filled, Coulomb interaction effects lead to a complicated multiplet structure. In a solid state environment the orbital degeneracy may be fully or partially lifted. In some cases, for example the cuprate high temperature superconductors,\cite{Science00} the degeneracy is fully lifted and the low energy physics may be described by the one-band Hubbard model in which the multiplet structure is trivial.\cite{Anderson86,Zhang88} However, for many materials of interest, including for example the  (La/Ca)TiO$_3$  series, the SrVO$_x$ materials, the (Sr/Ca)RuO$_3$ compounds and their Ruddlesden-Popper variants, the new Fe based superconductors and the A$_n$C$_{60}$ series of materials, as well as essentially all interesting lanthanide/actinide compounds, the orbital degeneracy is not fully lifted and nontrivial multiplet effects are expected to be important.  Of particular interest is the effect of orbital degeneracy on the Mott metal-insulator transition. It is generally believed \cite{Gunnarsson96} that the critical interaction strength required to drive a metal-insulator transition depends on the orbital degeneracy, being larger for systems with several degenerate orbitals than it is for one-orbital models. This gives rise to the physics of orbital selectivity, whereby a broken orbital symmetry, either spontaneous or induced by crystal symmetry, may drive some orbitals into insulating states. This phenomenon has for example been argued to be of crucial importance in understanding the insulating phase of LiTiO$_3$ \cite{Pavarini04} and of the metal insulator transition in Ca$_2$RuO$_4$.\cite{Moore07,Liebsch03}  

%ajmdec7
The dynamical mean field theory  (DMFT) provides a non-perturbative method to study the interplay of correlation effects and electron banding and has in particular produced insights into the correlation-driven (Mott) metal-insulator transition in the one-orbital model.\cite{Georges96} While the issue of the Mott transition in multiorbital systems has been addressed by various techniques, \cite{Slaverotor,Liebsch,deMedici08} a comprehensive picture has not emerged, in part because of the theoretical difficulties associated with the treatment of the various Hund and pair hopping terms required for a realistic treatment of partially filled $d$-orbitals.   Dynamical mean field theory maps a lattice problem onto a quantum impurity model (a finite size system coupled to a noninteracting bath of electrons) plus a self consistency condition. For systems in which the orbital degeneracy is fully lifted the quantum impurity model is a variant of the one-orbital ``Anderson Impurity Model", for which powerful numerical techniques have been known for many years. \cite{HirshFye1,Rubtsov05,CTHirshFye,ED1,NRG} However, these techniques encounter difficulties when applied to materials with partially filled, degenerate $d$-orbitals, where the on-site interaction includes both spin exchange and ``pair hopping" terms. The Hirsch-Fye method, which has been the standard approach for multi-orbital models with density-density interactions,  relies on a Hubbard-Stratonovich transformation of the interaction term. In the orbitally degenerate case the multiplicity of interactions requires many auxiliary fields, which become difficult to sample. Rotational invariance becomes very difficult to preserve and a severe sign problem is reported.\cite{Sakai} The proliferation of states also creates difficulties for exact diagonalization methods, although recent progress has been made along these line.\cite{Liebsch,deMedici08} 

In this paper we exploit  a recently developed \cite{Werner05,Werner06} impurity solver  which is free from the defects of the other methods. In this method the on-site Hamiltonian is solved exactly, and the coupling to the bath is treated by a perturbation expansion which is sampled stochastically via an importance-sampling procedure. The method allows a detailed and accurate treatment of thermodynamic quantities and (via analytical continuation) of dynamics, down to temperatures of the order of $0.1\%$ of the basic energy scales of the problem. An additional benefit of the method is that it provides information about which configurations of the correlated site make the dominant contributions to the partition function. While the computational effort of our method scales exponentially with the number of orbitals, it can easily handle three orbitals on desktop machines, and five orbitals on larger clusters. 
  
In this paper we use the method to analyze the ``three orbital" model which is relevant to materials such as LaTiO$_3$, SrVO$_3$ and  SrRuO$_3$, where the physics is dominated by electrons in the transition metal $t_{2g}$ orbitals. The model is also relevant to  electron-doped C$_{60}$, where the three orbitals correspond to the triplet of  LUMO states of C$_{60}$. We determine the metal-insulator phase diagram, study the response to perturbations which lift the orbital degeneracy and determine the orbital selectivity of the doped Mott insulating state. Our work builds on our previous investigation of a ``two-orbital" model \cite{Werner07Crystal} relevant to systems with $e_g$ symmetry. In the two orbital case, in the presence of strong Hund coupling, one has either a one-electron state or a filled (spin-polarized) shell. The new feature of the three orbital model is the case $n=2$, where one can have a  multielectron state with high local spin alignment but a partially filled shell.

\section{Formalism \label{Formalism}}

We study a model involving three orbitals (labeled by $a=1,2,3$),  with Hamiltonian
%ajmdec7 I wrote the band energy in a superficially more general way
%
\begin{equation}
H=\sum_{k, a,b,\sigma} \varepsilon^{ab}_k d^\dagger_{k,a,\sigma}d_{k,b,\sigma}-\sum_{i,a, \sigma}(\mu-\Delta_a) n_{i,a,\sigma}+\sum_i H_\text{loc}^i.
\label{Hdef}
\end{equation}
Here $i$ labels sites in a lattice and $k$ a wave vector in the first Brillouin zone, $n_{i,a,\sigma}=d^\dagger_{i,a,\sigma}d_{i,a,\sigma}$ is the density of electrons of spin $\sigma$ in orbital $a$ on site $i$, $\mu$ is the chemical potential, $\Delta_a$ is a level shift for orbital $a$ arising from a ligand field splitting and $\varepsilon^{ab}_k$ is the band  dispersion. For the following analysis, the relevant property of the dispersion  is the density of states $N(\omega)$ given by  ($\int (dk)$ symbolizes an integral over the appropriate Brillouin zone with the correct measure factors)
\begin{equation}
N^{ab}(\omega)%=N(\omega)\delta^{ab}
=\int (dk) \delta (\omega-\varepsilon^{ab}_k).
\label{dosdef}
\end{equation}
We have assumed that the symmetry is such that the local density of states is orbital-diagonal and  independent of $a$; this assumption holds for pseudocubic materials such as the La-titanates and the ``$113$'' Sr/Ca ruthenates, as well as for A$_n$C$_{60}$. We expect that the qualitative consequences of a symmetry breaking in the density of states are similar to those obtained by introducing an explicit crystal field splitting $\Delta_a$.

For the interaction term we take the standard Slater-Kanamori form (we have suppressed the site index)
\begin{align}
H_\text{loc}&=\sum_{a} U n_{a,\uparrow} n_{a,\downarrow}\nonumber\\
&+\sum_{a>b,\sigma} \Big[U' n_{a,\sigma} n_{b,-\sigma} +  (U'-J) n_{a,\sigma}n_{b,\sigma}\Big]\nonumber\\
&-\sum_{a\ne b}J(d^\dagger_{a,\downarrow}d^\dagger_{b,\uparrow}d_{b,\downarrow}d_{a,\uparrow}
+ d^\dagger_{b,\uparrow}d^\dagger_{b,\downarrow}d_{a,\uparrow}d_{a,\downarrow} + h.c.).
\label{H_delta}
\end{align}
Here $U$ is the intra-orbital and $U'$ the inter-orbital Coulomb interaction, while $J$ is the coefficient of the Hund coupling and pair-hopping terms. We adopt the conventional choice of parameters,  $U'=U-2J$, which follows from symmetry considerations for $d$-orbitals in free space and is also believed to hold in solids.  With this choice the Hamiltonian (\ref{H_delta}) is rotationally invariant in orbital space. The chemical potential required to obtain a given occupancy at fixed $U$ decreases as $J$ is increased; for example the condition for half filling is $\mu = \frac{5}{2}U-5J$.  We shall focus on the case $U>3J$,  in which (loosely speaking) the $U$ interaction controls the occupancy and once the occupancy is fixed the $J$ interactions then control the arrangement of the electrons among orbitals. 
%ajmdec7
For   $U<3J$ the physics is different: the local level first maximizes the spin and then adjusts the local occupancy accordingly. We are not aware of materials for which this regime is relevant.

We solve the model using the single-site dynamical mean field approximation,\cite{Georges96} which neglects the momentum dependence of the self-energy and reduces the original lattice problem to the self-consistent solution of a quantum impurity model given by the Hamiltonian
\begin{equation}
H_\text{QI} =-\sum_{a, \sigma}(\mu-\Delta_a) n_{a,\sigma} +H_\text{loc}+H_\text{hyb}+H_\text{bath}
\label{HQI}
\end{equation}
with 
\begin{eqnarray}
H_\text{hyb}&=&\sum_{k,a,\sigma} V_{k,a,\sigma}d^\dagger_{a,\sigma}c_{k,a,\sigma}+h.c., \\
H_\text{bath}&=&\sum_{k,a,\sigma}\varepsilon^\text{bath}_a(k)c^\dagger_{k,a,\sigma}c_{k,a,\sigma}.
\end{eqnarray}

The important quantity for the subsequent analysis is the hybridization function $\Delta^{a,\sigma}_\text{hyb}(\omega)$ which depends on orbital $a$, spin $\sigma$ and frequency and whose imaginary part is 
\begin{equation}
\text{Im}\Delta^{a,\sigma}_\text{hyb}(\omega)=\int (dk)\left|V_{k,a,\sigma}\right|^2\delta(\omega-\varepsilon^\text{bath}_a(k)).
\end{equation}
In the computations presented here we take an orbital-independent semi-circular density of states with band-width $4t$ (Bethe lattice). The $t_{2g}$ band widths for early-stage transition metal oxide compounds are of the order of $3eV$, so that $t\sim 0.75 eV$.

The hybridization function is fixed by a self-consistency condition \cite{Georges96} involving the impurity model Green's function $G^\text{QI}$, the self energy $\Sigma^\text{QI}$ of the quantum impurity model and the momentum integral of the Green's function  of the  lattice problem computed with $\Sigma^\text{QI}$,
\begin{equation}
G^\text{latt}_{a,\sigma}(i\omega_n)=\int d \varepsilon \frac{N^a(\varepsilon)}{i\omega_n-\varepsilon-\Sigma^\text{QI}_{a,\sigma}(i\omega_n) }.
\label{glocdef}
\end{equation}

We note that insulating solutions may easily be distinguished from metallic solution by the behavior of $G^\text{QI}(\tau)$: for an insulator at low $T$ this quantity drops exponentially as $\tau$ is increased from $0$ or decreased from $\beta$, while in a metallic phase the constant Fermi level density of states leads to a slow power-law decay. 

\begin{figure}[t]
\begin{center}
\includegraphics[angle=-90, width=0.9\columnwidth]{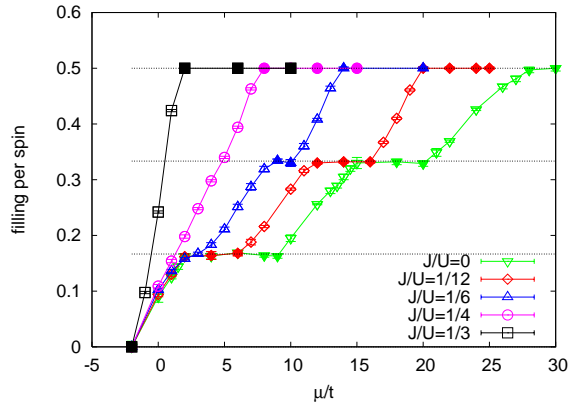}
\caption{(color online) Electron density $n$ per orbital per spin computed as a function of chemical potential $\mu$ for different values of the interaction parameter $J$ at $U/t=12$ and temperature $\beta t=50$.  The orbital symmetry of the Hamiltonian is unbroken ($\Delta_a=0$) and  orbital and spin symmetry were enforced in the calculation.  Plateaux in $n(\mu)$ correspond to Mott insulating states.  Open (full) symbols correspond to metallic (insulating) solutions. 
}
\label{nmu_u12}
\end{center}
\end{figure}

The simulations were performed using a continuous-time QMC solver which samples a diagrammatic expansion of the partition function in powers of the impurity-bath hybridization $H_\text{hyb}$. \cite{Werner05, Werner06} We monitored the particle densities in each orbital, the Green's functions and self energies of the impurity model, and the contributions of each eigenstate of $H_\text{loc}$ to the partition function. 
For a three-orbital model the dimension of the Hilbert space of $H_\text{loc}$ is $64$, so the series is constructed in terms of traces of products of $64 \times 64$ matrices combined with determinants made up of the the hybridization function $\Delta_\text{hyb}$ evaluated at different time arguments. The bottleneck of the simulation is the trace computation. To speed this up, it is important to group the eigenstates of $H_\text{loc}$ according to the conserved quantum numbers as explained in Ref.~\onlinecite{Haule07}. The matrix-representation of the operators $d$ and $d^\dagger$ then acquires a block structure, with blocks of size $\le 9$. This way, the simulation becomes efficient enough to run on a desktop machine. Our results were obtained using about 3-5 CPU hours per iteration.

\section{Metal-Insulator Phase Diagram; Orbitally symmetric case}

To map out the metal-insulator phase diagram we have computed the dependence of density (typically represented as a density per orbital per spin) as a function of chemical potential for various interaction values. Figure~\ref{nmu_u12} shows representative results. For sufficiently negative $\mu$ the solution we find has density $n=0$ (``band insulator"). As $\mu $ is increased, the density increases. For small $U$ the increase is smooth at all $\mu$, while at larger $U$   plateaux occur at which the density is fixed to the integer values $n=1,2,3$ (so the density per orbital per spin is fixed to $1/6,2/6,3/6$).  
%ajmdec7 minor rephrase
We identify the regions in which $n$ smoothly increases as metallic phases and the plateaux as Mott insulating regions; we have confirmed these identifications by examination of $G^\text{QI}(\tau)$. Metallic (insulating) solutions are plotted with open (full) symbols. At  the $U=12t$ value studied in Fig.~\ref{nmu_u12} we see that for $J/t=0$ and 1, we have plateaux at each integer $n$, for $J/t=2$ only at $n=2,3$ and for $J/t=3$ and 4 there is only a plateau at $n=3$. From similar traces at various values of $U,J$  we have constructed  metal-insulator phase diagrams in the plane of chemical potential $\mu$ and correlation strength $U$.   

\begin{figure}[t]
\begin{center}
\includegraphics[angle=-90, width=0.9\columnwidth]{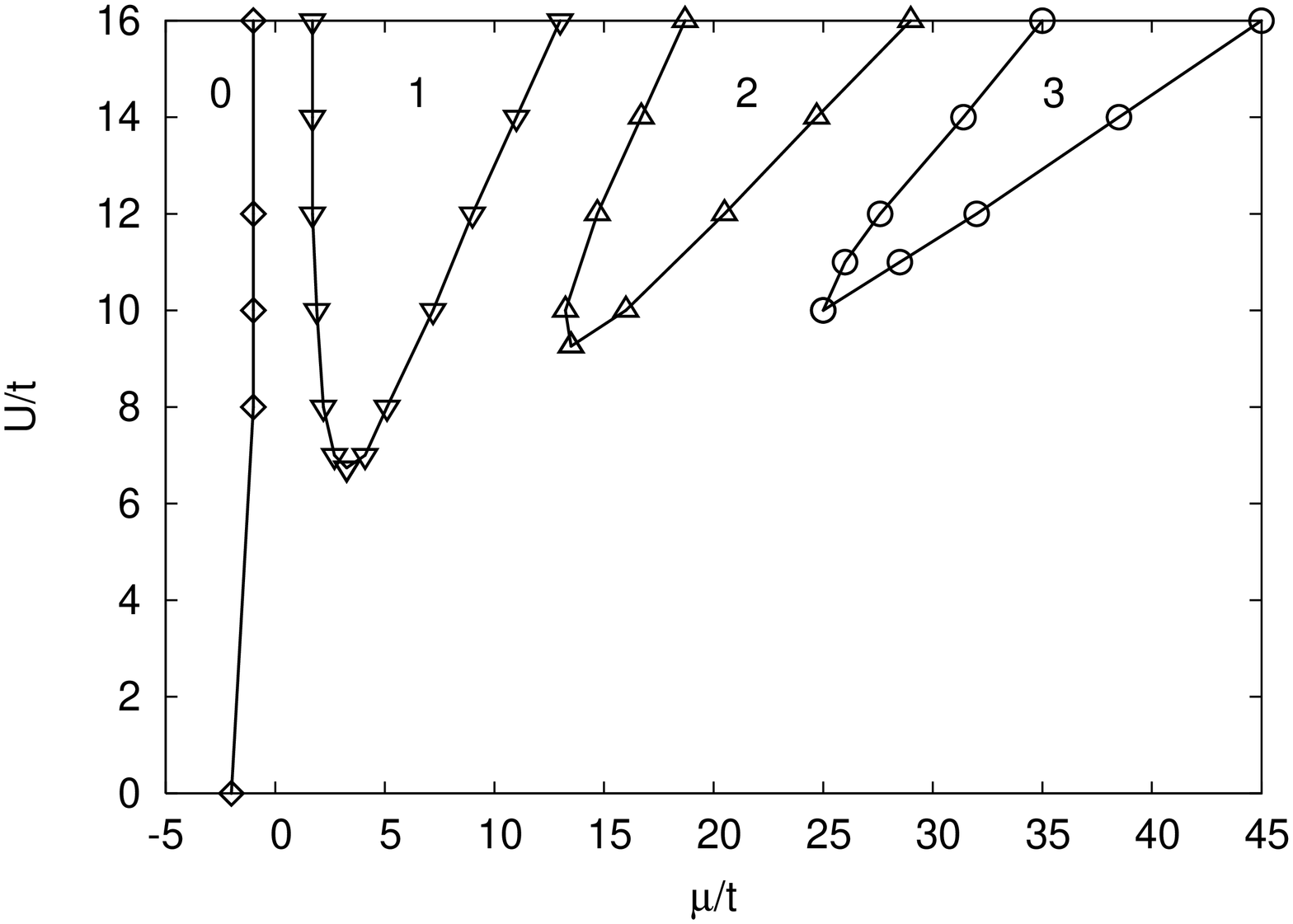}
\includegraphics[angle=-90, width=0.9\columnwidth]{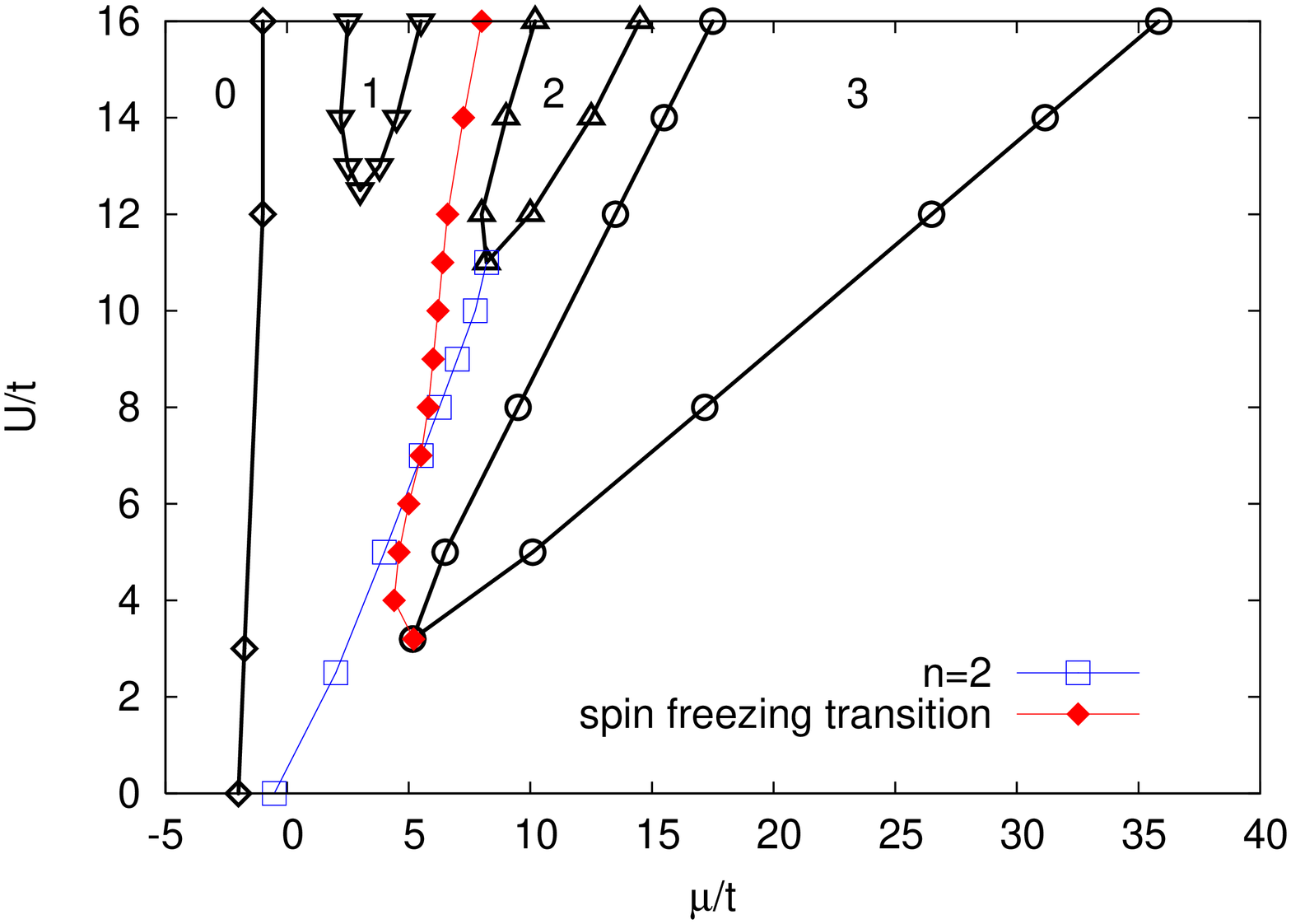}
\caption{Metal-insulator phase diagram presented in the space of chemical potential $\mu$ and interaction strength $U$ (measured in units of the quarter-bandwidth $t$) for $\Delta_a=0$,  $\beta t=50$ at Hund's coupling $J=0$ (upper panel) and $J=U/6$ (lower panel).  Orbital and spin symmetry were enforced in the calculation.  Error bars are of the order of the symbol size. The numerals in the lobes indicate the electron concentration per site in the insulating phases. In the lower panel the solid diamonds indicate the boundary of a spin freezing transition discussed in Ref.~\onlinecite{Werner08nfl}, while the line with squares plots the locus of $\mu$ and $U$ corresponding to the density $n=2$.
}
\label{lobe1}
\end{center}
\end{figure}

The upper panel of Fig.~\ref{lobe1} shows 
%ajmdec7 results
the phase diagram for $J=0$. We see that the critical $U$ required to drive a Mott transition depends weakly on density, ranging from $U=6.5t$ at $n=1$ to $U=10t$ at $n=3$. Positions and widths (in $\mu$) of the Mott lobes are only weakly dependent on band filling (at $U=16t$ the width is about $11t$ for all three lobes). The lower panel of Fig.~\ref{lobe1}  shows that the situation changes quite dramatically in the presence of a Hund coupling. The size of the 3-electron insulating lobe is substantially increased at the expense of the 2- and 1-electron lobes.  Furthermore, the value of $U_{c2}$ for the half-filled insulating state is reduced from $\approx 10t$ to $\approx 3t$, while the 2- and 1-electron  insulating lobes shift to higher values of $U$. 

Insight into the physics of the metal-insulator phase boundaries can be obtained by considering the atomic limit.  If $E_n$ denotes the lowest eigenvalue of the $n$-particle sector of $H_\text{loc}$, then an estimate of the Mott gap is
\begin{equation}
\Delta_\text{Mott}(n)=E_{n+1}+E_{n-1}-2E_n.
\label{Mottgap}
\end{equation}
The actual Mott gap is reduced by an amount of the  order of the bandwidth ($4t$) while the critical $U$ required to drive a metal-insulator transition may be estimated by comparing the strong coupling $\Delta_\text{Mott}$ to the  electronic kinetic energy $K=-\sum_{a,\sigma}\int (dk) \varepsilon_k \langle c^\dagger_{k,a,\sigma}c_{k,a,\sigma}\rangle$.

\begin{figure}[t]
\begin{center}
\includegraphics[angle=-90, width=0.9\columnwidth]{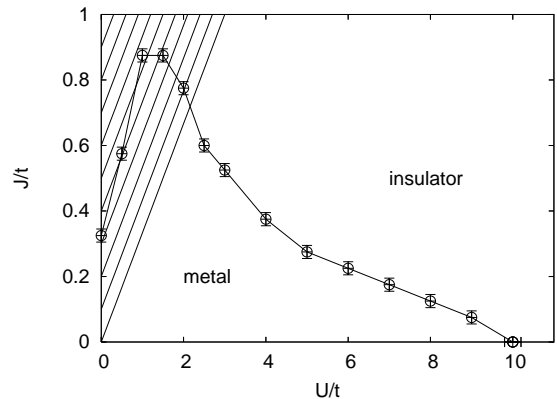}
\caption{Phase diagram in the plane of Slater-Kanamori parameters $U$ and $J$ calculated for the orbitally symmetric model ($\Delta_a=0$) at $\beta t=50$ and half filling ($n=3$). The hashed region ($U<3J$) corresponds to an effectively attractive Coulomb interaction; this situation does not normally occur in transition metal oxides and is not considered here.}
\label{phasediagram_halffilling}
\end{center}
\end{figure}

For $J=0$ the interaction term $H_\text{loc}$ may be rewritten 
%ajmdec7 important factor of 2 introduced
$H_\text{loc}=UN_\text{tot}(N_\text{tot}-1)/2$, so Eq.~(\ref{Mottgap}) gives $\Delta_\text{Mott}=U$ for all $n$. The upper panel of Fig.~\ref{lobe1} shows that at large $U$ the width in $\mu$ is almost the same for the three Mott lobes, consistent with this simple argument. Similarly the  $n$-dependence of the critical $U$ is consistent with the $n$ dependence of the noninteracting kinetic energy $K(n=1) \approx 1.4t$, $K(n=2)\approx 2.3t$ and $K(n=3)\approx 2.5t$. 

A non-zero $J$ term changes the energetics. The lowest energy state is of maximal spin and we find $\Delta_\text{Mott}(1)=\Delta_\text{Mott}(2)=U-3J$. However, for $n=3$, adding a fourth electron requires flipping a spin, so that $\Delta_\text{Mott}(3)=U+4J$.  For $J=U/6$ as in the lower panel of Fig.~\ref{lobe1} this becomes $\Delta_\text{Mott}(n=1,2)\approx U/2$  and $\Delta_\text{Mott}(3) \approx 5U/3$.  These considerations explain the comparable widths of the  Mott lobes for $n=1,2$  and the much larger width of the $n=3$ Mott lobe. The variation of the critical $U$ is more subtle. For $n=1,2$ the transition occurs at a sufficiently strong correlation that we may assume that each site is always in its maximal spin state, although our calculation is in the spin-disordered phase, so the direction of the moments is random from site to site.  The noninteracting kinetic energy should then be computed for fully spin polarized electrons, and should be reduced by a factor of $\sqrt{2}$ to account for the double-exchange physics of spin polarized electrons hopping in a paramagnetic environment. These considerations give $K(n=1)\approx K(n=2) \approx 0.8t$; the reduced $K$ and reduced $U$ account for the shift of the critical $U$.  For $n=3$, the situation is different: as $J$ becomes large, the constraint of total on-site spin polarization means that no low-energy states are available for conduction: there is only virtual hopping and as in the half-filled double exchange model one would have insulating behavior driven by $J$ only. This means that the kinetic energy is very rapidly suppressed by a non-vanishing $J$, explaining the rapid shift in the phase boundary seen in Fig.~\ref{lobe1} and in more detail in Fig.~\ref{phasediagram_halffilling}.

\begin{figure}[t]
\begin{center}
\includegraphics[angle=-90, width=0.9\columnwidth]{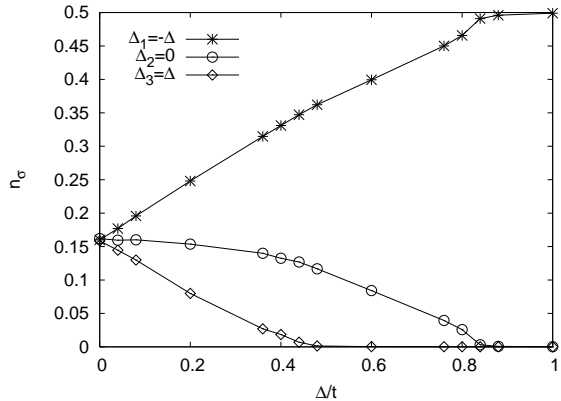}
\caption{Orbital filling as a function of crystal field splitting $\Delta$ in the symmetric case: $\Delta_1=-\Delta$, $\Delta_2=0$, $\Delta_3=\Delta$. The parameters are  $U/t=7<U_{c2}$, $J/U=1/6$,  $\beta t=50$, and the density at $\Delta=0$ corresponds to 1 electron. As the crystal field splitting is increased, band 1 (which is raised) empties out and undergoes a metal-band insulator transition near $\Delta/t\approx 0.4$. At the higher value $\Delta\approx 0.8t$ the second band empties out, leaving what is effectively a one orbital model for which $U>U_{c2}$ so the state is insulating. }
\label{1_1_1_splitting}
\end{center}
\end{figure}

\begin{figure}[t]
\begin{center}
\includegraphics[angle=-90, width=0.9\columnwidth]{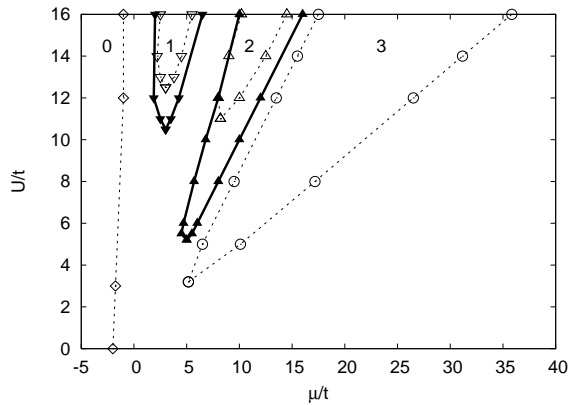}
\caption{Effect of a ``1 up, 2 down'' crystal field splitting on the 2-electron insulating phase. Heavy black line: boundary of the two electron Mott insulating state in the space of interaction $U$ and chemical potential $\mu$ computed for $\Delta_1=t$, $J=U/6$ and  $\beta t=50$. The crystal field  splits the threefold degenerate level into a doublet and a singlet, with the singlet lying higher.  Dashed lines: metal-insulator phase boundary for the same interaction parameters and $\Delta_a=0$ for comparison.
}
\label{lobe2}
\end{center}
\end{figure}

\section{Lifting of the Orbital Degeneracy: Metal-Insulator Phase Diagram and Mott Insulating States}

%ajm8/17
In this section we consider the consequences of an explicit breaking of the orbital symmetry of the model.  We focus mainly on $J>0$ and dopings between $n=1$ and $n=3$. The cases of $n=1$ and $n=3$ are straightforward. At $J>0$, the $n=3$ state is a filled shell, stable against orbital splitting for small differences among the $\Delta_a$, while for larger crystal field splitting a high-spin/low-spin transition will occur, with physics analogous to that discussed in the two orbital context in Ref.~[\onlinecite{Werner07Crystal}].  For $n=1$ the qualitative behavior is clear: the model becomes either an effective one orbital model or an effective two orbital model; the physics of these two cases has been previously discussed.\cite{Georges96,Werner07Crystal} As an example we show in Fig.~\ref{1_1_1_splitting} the  evolution of the orbital occupancy under a ``trigonal'' crystal field splitting $\Delta_1=\Delta$, $\Delta_2=0$, $\Delta_3=-\Delta$ which separates all three orbitals. We see that as the splitting is increased one band becomes depopulated and then, at a higher $\Delta$, the second band empties out, leaving  an orbitally polarized Mott insulator.  This behavior is consistent with the proposal of Pavarini {\it et al.},\cite{Pavarini04} who argued that the insulating behavior of the $n=1$ material LaTiO$_3$ is due to a relatively strong ligand field which splits the degeneracy of the three orbitals.

%ajmdec7 rewording
We focus now on the case $n=2$, which is relevant for example to  SrRuO$_3$  and the Ruddlesden-Popper materials Sr$_{n+1}$Ru$_n$O$_{3n+1}$. In studying these cases our main focus is on the simplest symmetry breaking, a cubic-tetragonal distortion which splits the $3$-fold degeneracy of the $t_{2g}$ state into a singlet and a doublet. We parametrize this splitting by moving one orbital (which we take to be ``orbital 1" by an energy $\Delta$ while keeping the other two fixed, so $\Delta_1=\Delta$, $\Delta_2=\Delta_3=0$. There are two cases: either the doublet lies lower than the singlet (``1 up, 2 down'', $\Delta>0$) or the reverse (``1 down, 2 up'', $\Delta<0$).  While we implement here the symmetry breaking  by shifting the orbital energies, other ways of breaking the symmetry (e.g. choosing different bandwidths) will have similar effects.  

We begin by considering the large-$U$, Mott insulating regime. Figure~\ref{lobe2} compares the metal insulator phase boundary computed for the orbitally symmetric model to the location of the $n=1$, 2 Mott lobes computed for a ``1 up, 2 down" crystal field $\Delta_1$ of magnitude $t$. Magnetic and orbital ordering are suppressed. Lifting the orbital degeneracy is seen to have a very substantial effect on the $n=2$ Mott phase and a noticeable but less dramatic effect on the $n=1$ Mott lobe. The critical interaction strength needed to drive the two electron phase insulating is seen to be reduced to less than half of the value found in the orbitally symmetric model. The width of the 2-electron insulating plateau is enhanced, but to a lesser extent: the increase in the width is approximately $\Delta_1$. Both positive (1 up, 2 down) and negative (1 down, 2 up) crystal field splittings stabilize the insulator, but the effect of a positive $\Delta_1$ (which shifts band 1 up) is much larger.  For $\Delta_1=-t$ (not shown) the end point of the 2-electron lobe is $U_{c2}\approx 9.7t$.   The difference occurs because if one level is shifted up, the $n=2$ electron state effectively becomes a filled shell which (as can be seen for the 3 electron state in Fig.~\ref{lobe1}) is particularly stable.

\begin{figure}[t]
\begin{center}
\includegraphics[angle=-90, width=0.9\columnwidth]{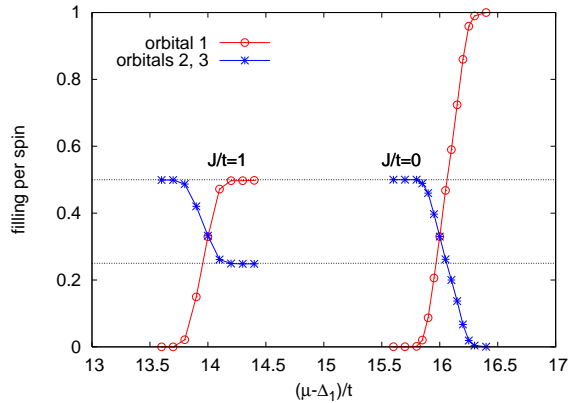}
\caption{(color online) Orbital filling as function of crystal field splitting $\Delta_1$ computed for the two electron insulating state with  $U/t=12$, $\beta t=50$ and indicated values of $J/t$. In order to display all of the curves on the same figure the crystal field coordinate is chosen to be $\mu-\Delta_1$ where the value of $\mu$ corresponds to $\mu/t$=16, 14 for $J/t$=0, 1, respectively. The ground state is insulating for all points shown. The red lines with circles correspond to the occupancy of orbital $1$ and the blue lines with stars to the occupancy of orbitals $2,3$. The conventions are such that increasing  $\Delta_1$ to positive values (moving to the left on the plot) shifts the non-degenerate orbital up (decreasing its occupancy). The offset between the curves for  different $J$ values arises because of the $J$-dependence of the location of the Mott lobes. Dashed horizontal lines are shown at filling 1/4 and 1/2.  
%ajmdec7 The upper panel shows the result if orbital order is suppressed, the lower panel if orbital order is allowed. In an orbitally ordered state, the occupancy in a given orbital oscillates between the two values shown in the figure, which should be viewed as the occupancies in sublattice $A$ and $B$.
%Orbital filling as function of crystal field splitting $\Delta_1$ computed for the two electron insulating state with  $U/t=12$, $\beta t=50$ and indicated values of $J/t$. In order to display all of the curves on the same figure the crystal field coordinate is chosen to be $\mu-\Delta_1$ where the value of $\mu$ corresponds to $\mu/t$=16, 14, 10 for $J/t$=0, 1, 2, respectively. The ground state is insulating for all points shown. The lines with circles correspond to the occupancy of orbital $1$ and the lines with stars to the occupancy of orbitals $2,3$. The conventions are such that increasing  $\Delta_1$ to positive values (moving to the left on the plot) shifts the non-degenerate orbital up (decreasing its occupancy). The offset between the curves for  different $J$ values arises because of the $J$-dependence of the location of the Mott lobes. Dashed horizontal lines are shown at filling 1/6, 1/4, 1/3, and 1/2.  
}
\label{polarizationall}
\end{center}
\end{figure}

Figure~\ref{polarizationall} presents  the response of the 2-electron insulating state to crystal field splitting for two values of $J$. 
%ajmdec7 The upper panel shows the results with suppressed orbital ordering (averaged Green's functions in orbital 2 and 3), while the lower panel shows the result obtained with averaging over spin, but not over orbitals. In regions where oscillating solutions are found, we plot both occupancies, which should be interpreted as orbital occupancies on two sublattices. 
In the high-spin filled shell case of two electrons in two orbitals studied in Ref.~\onlinecite{Werner07Crystal}, the insulating state (for $J>0$) did not respond at all to a weak crystal field splitting. Here, because at density $n=2$  the $\Delta_a=0$ state is not a filled shell, the two electron insulating state responds even to an infinitesimal crystal field splitting: the orbital susceptibility is non-vanishing. 

The value of the Hund coupling $J$ has important effects on the response to a crystal field. At $J=0$ (rightmost traces in Fig.~\ref{polarizationall} there is no energetic barrier to placing two electrons in the same orbital. If $\Delta_1$ is decreased ($\mu-\Delta_1$ increased) the occupancy of band 1 increases to 1 per spin while the occupancy of the other two bands decreases smoothly to zero. If $\Delta_1$ is increased, the non-degenerate state empties out while the occupancy of the two degenerate states remains equal, and approaches $1/2$ per spin per orbital. 
%ajmdec7 From the lower panel we see that the $J=0$ insulating state is orbitally ordered everywhere except for large negative crystal field splitting (filled shell).
If $J>0$ (left hand traces) the situation changes: at $\Delta_a=0$ the lowest energy state is spin triplet, so that as $\Delta_1$ is decreased only $1/2$  electron per spin  can populate orbital 1, which leads to an average distribution of $(1/2, 1/4, 1/4)$.  For even larger $\Delta_1$ a high-spin/low spin transition will occur, but we do not consider this here.
Depending on the degree of band-nesting, the states considered here may become unstable to orbital ordering. Magnetic and orbital ordering will be discussed in a future paper.\cite{ChrisChan08}

\begin{figure}[t]
\begin{center}
\includegraphics[angle=-90, width=0.9\columnwidth]{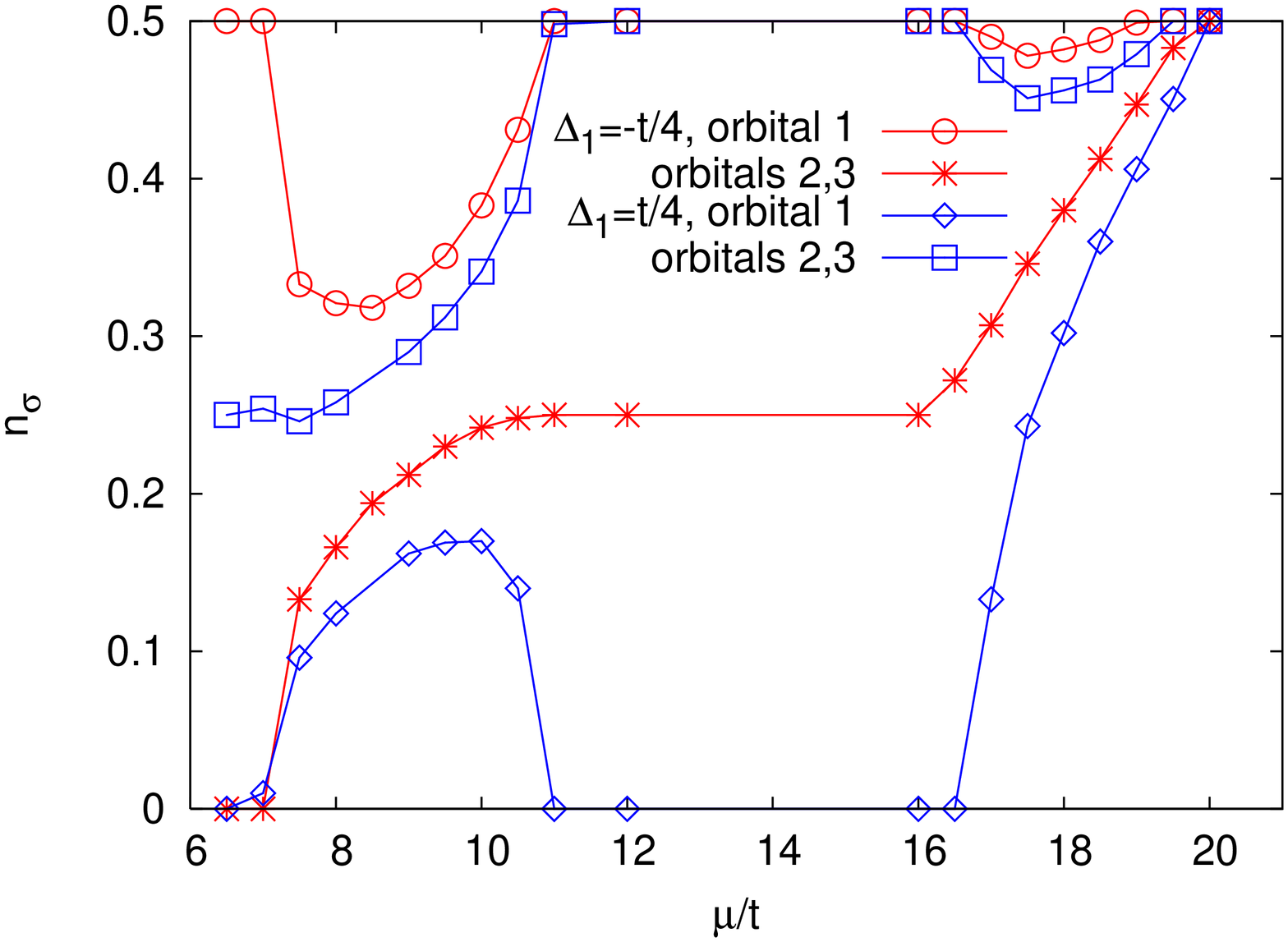}
\includegraphics[angle=-90, width=0.9\columnwidth]{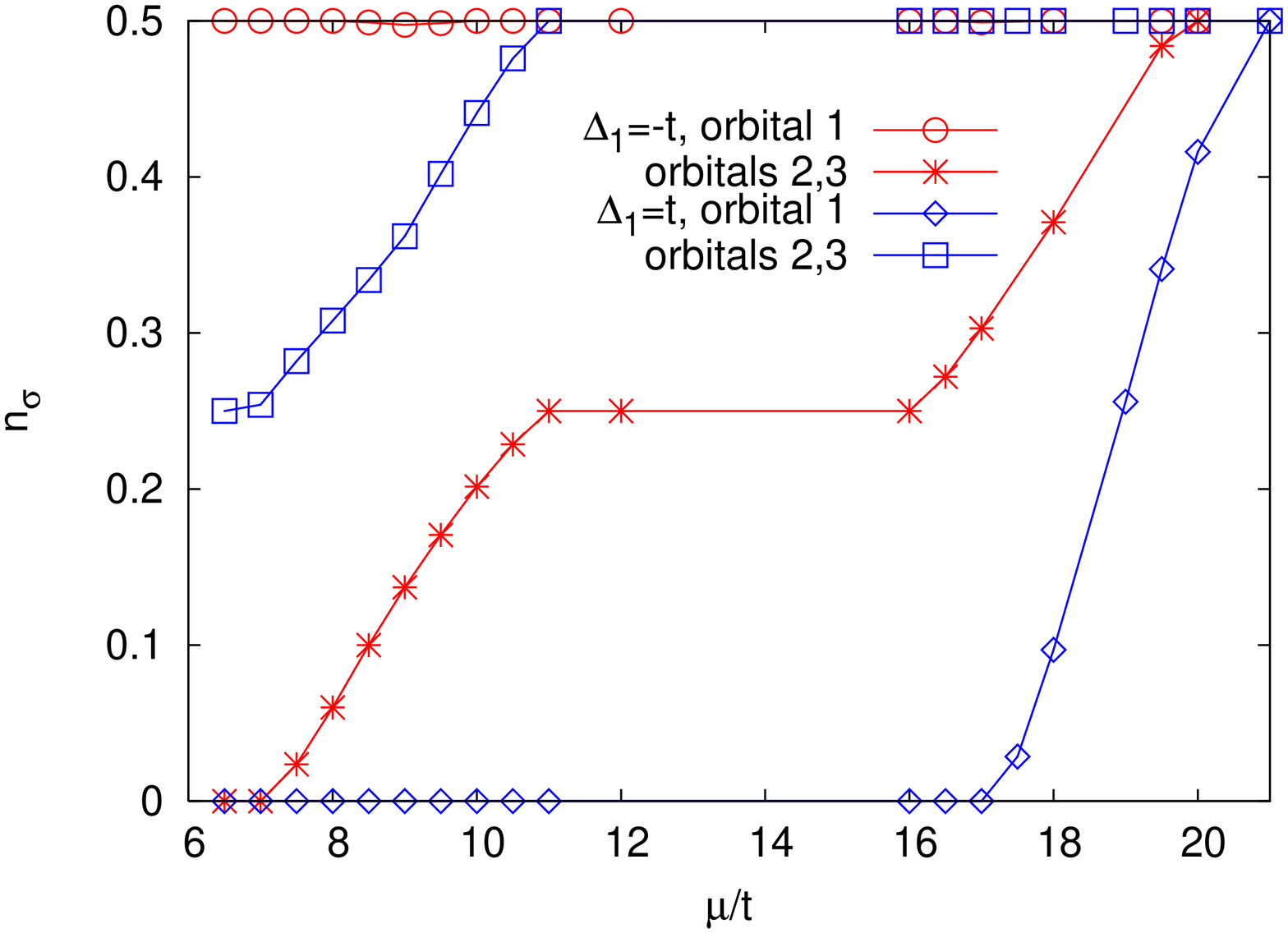}
\caption{$\mu$-dependence of the orbital occupancy per spin  $n_\sigma(\mu)$ for $U/t=12$, $J/t=1$ at $\beta t=50$ and in the presence of a cubic-tetragonal crystal field $\Delta_1=\pm 0.25t$ (top) and $\Delta_1=\pm t$  (bottom). The crystal field splits the threefold degenerate $d$ level into a doublet and a singlet, with the doublet lying higher or lower according to the sign of $\Delta_1$. The singlet orbital is labeled as ``orbital 1" and is denoted by open circles (red on-line) for $\Delta_1<0$ or  diamonds (blue on line) for $\Delta_1>0$; the doublet orbitals are labeled as orbitals 2,3 and are denoted by stars (red on-line) for $\Delta<0$ or squares  (blue on line) for $\Delta>0$.  Magnetic ordering was suppressed by averaging the Green function over spin  and additional ordering of orbitals 2 and 3 was suppressed by averaging the Green functions in orbitals 2 and 3. The chemical potential range runs from the $n=1$ Mott phase ($\mu \sim 6t$) to the $n=3$ Mott phase ($\mu \sim 20t$). Insulating phases are visible as plateaux in all three densities and occur only at integer total density. Orbitally selective Mott phases are visible as plateaux in one density with the other(s) varying with $\mu$. }
\label{n_mu_delta}
\end{center}
\end{figure}

%ajmdec7 This distribution is seen to be unstable to an additional, spontaneous orbital order which leads to a disproportionation between the occupancies of the two minority orbitals and thus an occupancy of $(1/2, 1/4\pm 1/4, 1/4\mp 1/4)$. %For sufficiently large $J$ in fact the disproportionation occurs even at $\Delta_a=0$.  
%In fact, for $J>0$ all insulating states, except those with large positive crystal field splitting are orbitally ordered. 
%ajm8/17At large values of $J$ one finds oscillating solutions with fillings $1/4\mp \delta$ in bands 2 and 3, which indicates a tendency towards orbital ordering.  
%However we see that at both $J=t$ and $J=2t$  
%the $(1/2, 1/4, 1/4)$ is unstable towards orbital ordering (seen as a difference in the occupancy of the two higher-lying bands. At $J=2t$ the difference is manifest even without crystal field splitting, indicating a spontaneous  orbital ordering instability within the Mott phase.THIS IS INTERESTING; IS THERE ANYTHING MORE WE WANT TO SAY ABOUT THIS
%Figure~\ref{polarizationall} shows the response of the 2/6 insulating phase to a crystal field splitting (shift of band 1). If band 1 is shifted up ($\Delta_1>0$), the filling in the 3 bands evolves from $(1/3,1/3,1/3)$ to $(0, 1/2, 1/2)$. If band 1 is shifted down ($\Delta_1<0$) one ends up in an insulating state with filling $(1,0,0)$ if $J=0$, or $(1/2, 1/4, 1/4)$ if $J>0$. In the latter case, we find oscillating solutions with fillings $1/4\mp \delta$ in bands 2 and 3, which indicates a tendency towards orbital ordering.  

\section{Crystal Fields and the Doped Mott Insulator}

We now consider the behavior occurring as the $n=2$ Mott insulator is doped in the presence of a non-vanishing crystal field splitting. 
%We suppress orbital order by averaging over orbitals 2 and 3 (the doped metallic states, in contrast to the insulating states, are not susceptible to orbital ordering). 
The results presented in this section pertain to an orbitally symmetric solution, and may be changed if orbital order occurs. Our preliminary results are that except very close to the Mott insulating phase boundaries, the doped states are stable against staggered orbital ordering.\cite{ChrisChan08} %Orbital-order related  issues will be discussed more detail elsewhere \cite{ChrisChan08}. 

Representative data are shown in Fig.~\ref{n_mu_delta} which plots the dependence of orbital occupancy on chemical potential for a relatively small ($|\Delta_1|=0.25t$, upper panel) and relatively large ($|\Delta_1|=t$, lower panel) magnitude of the crystal field splitting. Results for both ``1 down, 2 up" (negative $\Delta_1$) and ``1 up, 2 down" (positive $\Delta_1$) crystal field splittings are shown.  The chemical potential range covers dopings  from the $n=1$ to the $n=3$ Mott insulating state.  The interaction parameters $J=t$ and $U=12t$ are such that the model is insulating at all three of the integer fillings $n=1,2,3$.

To discuss the figure it is convenient to begin with  the ``1 down, 2 up", $\Delta_1<0$ case (circles and stars, red on-line) and to discuss the behavior as the $n=1$ Mott insulating state found at  $\mu \sim 6t$ is doped. Although the orbital susceptibility of the $n=1$ Mott insulating state is finite, even the weaker of the two crystal field splittings shown here is larger than the ``orbital superexchange" and leads to complete orbital polarization. The favored orbital is fully occupied (density $n=0.5/\text{spin}$) and the disfavored orbitals are empty. Now consider adding electrons to the ``1 down, 2 up" state. In the weak crystal field case (upper panel) we see that  (within our resolution) the doping-driven Mott transition out of the $n=1$ state is first order: the  orbital polarization drops dramatically on doping so that in addition to adding electrons, doping leads to a transfer of electrons from the highly occupied to the less highly occupied orbital. The resulting ``orbitally polarized Fermi liquid" state evolves smoothly upon doping to the obvious two-electron Mott state, characterized by the expected $1/2,1/4,1/4$ occupancy per spin. %(we average here the occupancy of orbitals 2 and 3). 
As electrons are added to this two electron state, we find a small orbitally selective Mott region with band 1 still insulating but bands 2 and 3 metallic. At larger chemical potential an insulator-metal transition takes place in band 1, leading to an initial decrease of the orbital polarization. This state evolves smoothly to the three electron, orbitally symmetric state.  Thus for small crystal field splitting, ``orbitally selective Mott behavior" only occurs very close to the insulating concentrations. Consider now the larger crystal field splitting (lower panel). In this case the doped state is in the orbitally selective Mott phase: at all chemical potentials the orbital favored by the crystal field splitting remains at the Mott occupancy of $n=1$ and the carrier density varies only in the disfavored orbital, so that one has effectively a model of two bands of carriers coupled to a spin-$1/2$ arising from the filled orbital.  Similar effects were also noticed recently by Liebsch \cite{Liebsch07chargetransfer} in a study of La$_{1-x}$Sr$_x$TiO$_3$ that corresponds to our model in the range $(0<n<1)$.

We next turn to the ``1 up, 2 down" $\Delta_1>0$ case (squares and diamonds, blue on-line). At $n=1$ we see again that the crystal field splitting is large enough to fully polarize the Mott insulator ($0, 1/4,1/4$ orbital occupancy). %We suppress any possible orbital ordering by averaging over the occupancies of the majority orbitals, so the state is characterized by a $0, 1/4,1/4$ orbital occupancy.  
In this case, at weak crystal field splitting, the occupancy of the majority orbitals increases smoothly with doping (almost all dopants go into the initially empty band, leading to a jump at the metal-insulator transition). At larger chemical potential, there is an abrupt transition to the $n=2$ Mott phase with $0, 1/2,1/2$ orbital occupancy. On further doping to the range $2<n<3$ we observe phenomena analogous to those found on doping the $n=1$ ``1 down, 2 up" state: doping leads to a charge transfer between orbitals which reduces the degree of orbital disproportionation. At the larger crystal field splitting the behavior is different. Between $n=1$ and $n=2$ the minority orbital remains empty; the crystal field splitting is large enough to make the material effectively a two orbital band insulator. Between $n=2$ and $n=3$ the state is an orbitally selective Mott state, with the minority band partially occupied and coupled to the spin-1 formed by the majority states.  This physics has also been discussed very recently in Ref.~\onlinecite{deMedici08}.

\begin{figure}[t]
\begin{center}
\includegraphics[angle=-90, width=0.9\columnwidth]{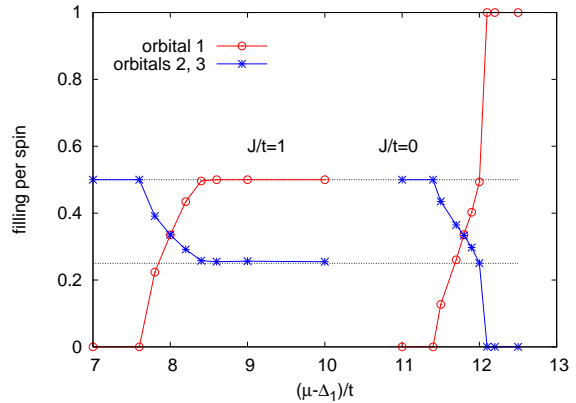}
\caption{ (color online) Orbital filling as function of crystal field splitting computed for the two electron  state with  $U/t=8$, $\beta t=50$ and indicated values of $J/t$. In order to display all of the curves on the same figure the crystal field coordinate is chosen to be $\mu-\Delta_1$. The red lines with circles correspond to the occupancy of orbital $1$ and the blue lines with stars to the occupancy of orbitals $2,3$. 
%ajmdec 7 The top panel shows the result with suppressed orbital ordering. 
While a metal-insulator transition is evident in the curves for $J/t=0$ and those for $J/t=1, \Delta_1>0$, the $J/t=1, \Delta_1<0$ curves exhibit a transition to an orbital selective Mott state (band 1 insulating, bands 2 and 3 metallic). 
%The lower panel shows the results with orbital ordering. % (the two values of the blue curve with stars correspond to the occupancy on sublattice A and B). The orbital selective Mott phase is replaced by an orbitally ordered insulator. 
%Orbital filling as function of crystal field splitting computed for the two electron  state with  $U/t=8$, $\beta t=50$ and indicated values of $J/t$. In order to display all of the curves on the same figure the crystal field coordinate is chosen to be $\mu-\Delta_1$ where for each $J$ the value of $\mu$ corresponds to an occupancy of $n=2$.  The lines with circles correspond to the occupancy of orbital $1$ and the lines with stars to the occupancy of orbitals $2,3$. The conventions are such that increasing  $\Delta_1$ to positive values (moving to the left on the plot) shifts the singlet up, decreasing its occupancy.  
}
\label{polarizationu400}
\end{center}
\end{figure}

\section{Crystal Fields in the Metallic State}

This section considers the effect of crystal field splitting for weaker interactions $U<U_{c2}$ where at $\Delta_a=0$ the system is in the metallic phase. In Fig.~\ref{polarizationu400} we plot the variation in orbital occupancies as the crystal field is varied at fixed $\mu$ corresponding to $n=2$, as was done in Fig.~\ref{polarizationall} for a stronger $U$. 
%ajmdec7
The figure shows results obtained by averaging the Green's functions of orbitals 2 and 3. %, while the lower panel shows the results in the presence of orbital ordering. 
For the $U=8t$ considered here, the $\Delta_1=0$ metallic phase is characterized by an orbital susceptibility $\chi_\text{orb}=-\frac{d(n_1-(n_2+n_3)/2)}{d\Delta_1}$ with some $J$ dependence but a typical magnitude of  $\sim 0.2-0.3/t$.   As $\Delta_1$ is increased the disfavored orbital 1 empties out and the occupancy of the favored orbitals increases. At $J=0$ (right hand side of the figure)  we see that in the ``1 down, 2 up" case, an apparently first order transition to a (1,0,0) insulating state occurs as the magnitude of the crystal field splitting increases, whereas in the ``1 up, 2 down" case a transition occurs to Mott state with two electrons in two orbitals. %Only states very close to the metal-insulator transition are susceptible to orbital ordering. 
At $J=0$ all possible ways of arranging the two electrons among the two orbitals are degenerate; the degeneracy would be lifted by intersite effects. 

In the more physically relevant $J>0$ case  (left hand side of figure)  a crystal field splitting of the ``1 down, 2 up" type leads to an orbitally selective Mott state. 
%ajmdec7  if orbital ordering is suppressed, or a Mott insulating state with a further orbital disproportionation in the minority orbitals if orbital ordering is allowed. The latter is the $(1/2, 1/4\pm 1/4, 1/4\mp 1/4)$ state which is also evident in Fig.~\ref{polarizationall}. %Examination of the Green's functions for the minority orbitals indicates that this state is metallic. IS THIS RIGHT?? -- no, they are in fact all insulating
On the other hand, increasing the magnitude of a crystal field splitting of the  ``1 up, 2 down" type induces a transition to a $(0, 1/2, 1/2)$ insulating state. % which is not orbitally ordered. 
Again, the computations presented here are for an orbitally disordered Mott state. %The effects of orbital ordering will be considered elsewhere \cite{ChrisChan08}
In particular the orbital selective Mott phase would be susceptible to orbital ordering.

%ajmdec7 We see from this figure that orbital ordering can transform an orbital selective Mott state (which is metallic in bands 2, 3) into an insulating state. Thus, in the presence of orbital ordering, the insulating lobes shown in Fig.~\ref{lobe2} would grow at the expense of the orbital selective Mott region.

\begin{figure}[t]
\begin{center}
\includegraphics[angle=-90, width=0.9\columnwidth]{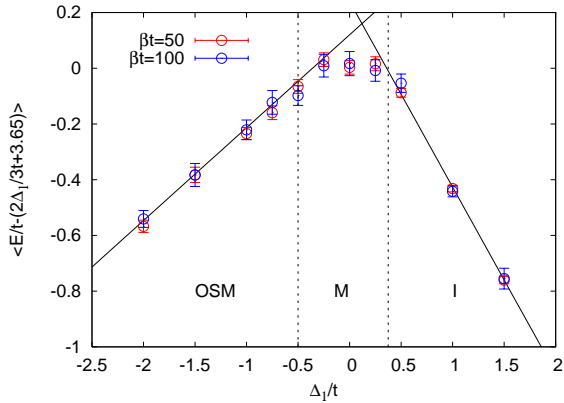}
\caption{Average energy $E$ of the 2/6 filled state at $U/t=8$, $J=U/6$ as function of crystal field splitting $\Delta_1$ measured relative to the energy of the $\Delta_1=0$ state. To compensate the asymmetry produced by raising/lowering one orbital while leaving the other two in place, we have subtracted $2\Delta_1/3t$. The results with $\Delta_1/t\lesssim-0.5$ are in an orbital selective Mott state with the lower orbital insulating and the other two metallic, those in the range $-0.5\lesssim \Delta/t < 0.5$ are metallic in all bands and the solutions at $\ge 0.5$ are insulating. In these phases we observe a linear behavior with slopes 1/3 and -2/3.}
\label{e_delta}
\end{center}
\end{figure}

Figure~\ref{e_delta} shows the energetics of the orbitally selective Mott transition %(with orbital ordering suppressed) 
at $U=8t$ and $J\approx 1.2t$ with density $n=2$. The chemical potential has been adjusted to keep the occupancy fixed. 
We see that in the metallic phase the energy is hardly affected, while in the orbitally selective and insulating phases the energy drops linearly with $\Delta$, with a coefficient given by the occupancy of the filled orbital. These energetics are important because in several materials (including for example Ca$_2$RuO$_4$) the metal insulator transition is of the orbitally selective type and is accompanied by a lattice distortion which acts to increase the crystal field splitting. 
%The orbital susceptibility (variation of difference in orbital occupancies with $\Delta^{CF}$) susceptibility remains approximately constant in the entire metallic range I AM NOT SURE THAT THIS IS TRUE. WE SEE BIG CHANGES AS YOU GET NEAR THE MOTT BOUNDARIES.. 

\section{Comparison to Experiment}

In this section we place a few relevant materials on our calculated phase diagram and discuss implications of our results. We begin with SrVO$_3$, a pseudocubic material characterized by 1 electron in the $t_{2g}$ shell, a bandwidth corresponding to $t\approx 0.7eV$, a $U \approx 5eV \approx 7t$ and $J\approx U/7$.\cite{Pavarini04} SrVO$_3$ is a good metal, with a modest mass enhancement; it is not believed to be close to the Mott transition. This behavior is consistent with our phase diagram: the value of $J/U$ is similar to that used to construct the lower panel of Fig.~\ref{lobe1} and $U=7t$ is quite far from the $n=1$ Mott lobe. The metallic behavior is seen to be a consequence of the non-vanishing value of the Hund coupling $J$. Without $J$, the material would be very close to the Mott transition.  In the related material LaVO$_3$ \cite{Mikokawa96,Sawada96,Fujioka08}  the change Sr $\rightarrow$ La implies that the $d$ shell filing changes from $1$ to $2$. The material also exhibits a moderate orthorhombic distortion away from cubic symmetry, of the ``1 down, 2 up" type.  Using the $U$ values  obtained from the singly-occupied system we find that the materials would be metallic (albeit in the spin-frozen phase discussed in Ref.~\onlinecite{Werner08nfl}).

We next consider LaTiO$_3$, in which the bandwidth is such that our parameter $t\sim 0.5-0.7eV$ and $U \sim 4-5eV$ $\sim 5-10t$, with $J \sim U/6$. Here examination of the phase diagram reveals that within the single site dynamical mean field theory, and in the absence of orbital ordering, the material is not predicted to be a Mott insulator. However, it is now known that in the material a substantial local trigonal distortion occurs.\cite{Cwik03,Hemberger03} Our results lend support to the idea, advanced in previous papers,\cite{Pavarini04,Okatov05} that the trigonal distortion is essential to the insulating behavior. The trigonal distortion, by lowering one orbital, will effectively convert the problem into a one orbital model. Figure~\ref{1_1_1_splitting} shows that the amplitude of the distortion must be large, providing a level splitting of the order of one quarter of the bandwidth. One difficulty with this interpretation is that the insulating gap in LaTiO$_3$ is only about $0.2eV$, whereas in a single-orbital Mott insulator the gap is of order $U-2t\sim 2eV$. The small value of the gap suggests that effects beyond the scope of the single site dynamical mean field theory are important in this material. 

We now to turn to the Sr/Ca ruthenates. These are materials with two holes in the $t_{2g}$ shell; after a particle-hole transformation they map on to the two electron case of the model studied here. The ``113'' materials (Sr$_{1-x}$Ca$_x$)RuO$_3$ crystallize in an orthorhombic structure slightly distorted from the cubic perovskite form. Both the Sr and Ca materials are metallic, with the Sr being ferromagnetic below $T_c \approx 150K$. The $t_{2g}$ bands have a bandwidth of approximately $2.5eV$ \cite{Mazin97} corresponding to a $t$ parameter of about $0.6eV$ in the notations of the present paper. The $U$ and $J$ values are not established for these compounds but must be substantially less than the $U=10t \sim 6eV$ needed to drive a Mott transition. 

The ruthenates also exist in a layered form (Sr/Ca)$_2$RuO$_4$. Here the electron counting is the same as in the ``113'' ruthenates but the tetragonal crystal structure means that two of the orbitals have an effectively one dimensional dispersion and the remaining one has a two dimensional dispersion. Thus, a substantial anisotropy is already present in the band structure even in the absence of explicit ligand field splitting. At room temperature all members of the Sr/Ca series are metallic, but  as temperature is reduced the Ca material undergoes a transition to an insulating phase, accompanied by a lattice distortion \cite{Jung03}  while Sr$_2$RuO$_4$ remains metallic to lowest temperatures. In a recent Letter \cite{Liebsch07}  Liebsch and Ishida have (in agreement with the prior proposals of Jung {\it et al.})\cite{Jung03} argued that the insulating phase should be understood as a consequence of a ``1 up, 2 down'' crystal field.  Figure~\ref{lobe2} shows that a moderate crystal field splitting of the ``1-up, 2-down" type can substantially decrease the critical $U$ required to drive a metal-insulator transition at $n=2$.  In this interpretation, the  effective crystal field is small in Sr$_2$RuO$_4$ but increases with Ca-doping, driving a metal-insulator transition analogous to that shown on the left hand side of  Fig.~\ref{polarizationu400}.  The argument in favor of a small effective crystal field splitting in Sr$_2$RuO$_4$ is the near equality of the occupancies of the $d_{xy}$, $d_{xz}$ and $d_{yz}$ orbitals. A previous weak coupling, Hartree-Fock analysis by one of us and Okamoto \cite{Okamoto04} found that at small $U$ a non-vanishing $J$ stabilized the systems against crystal field distortions. On the basis of this calculation it was argued that the near degeneracy of orbital occupancies in Sr$_2$RuO$_4$ despite the highly anisotropic crystal structure was an interaction effect. Figure~\ref{polarizationu400}  and the results of Ref.~\onlinecite{Liebsch07} suggest that the results of Ref.~\onlinecite{Okamoto04} do not survive beyond the weak coupling limit, so that the near equality of orbital occupations in Sr$_2$RuO$_4$ must be regarded as accidental, with the asymmetry of the bandwidths and of the crystal field levels compensating one another to leave a small effective splitting. If the near vanishing of effective crystal field splitting in Sr$_2$RuO$_4$ is accidental then it is very reasonable that the changes induced by Ca substitution can move the system away from the accidental degeneracy.

An issue with this interpretation is that at higher temperatures $T \gtrsim 350K$ Ca$_2$RuO$_4$ is metallic, while the standard single-site dynamical mean field theory predicts that the material should become more insulating as the temperature is raised. We suggest, following Jung {\it et al.},  that the temperature-driven  first order transition can be understood in terms of a temperature-driven lattice distortion. Indeed the energy diagram (Fig.~\ref{e_delta}) in combination with a reasonable free energy for lattice distortions,  implies a first order transition.  From this figure we see that the energy gain per orbital $\Delta E$ from a ``1 up, 2 down" distortion produced by a crystal field $\Delta$ is approximately $\Delta E=\Delta-\Delta_0$ for $\Delta>\Delta_0$ with   $\Delta_0$ a fraction of the bandwidth parameter $t$. Let us suppose that  a lattice distortion of the type observed by Ref.~\onlinecite{Jung03} produces a crystal field splitting $\Delta$  and that the free energy cost of this distortion is $\Delta F=\Delta^2/({2\bar \Delta}(T))$ with ${\bar \Delta}(T)$ a temperature dependent free energy scale which increases as $T$ is decreased, representing the entropy of thermal phonon fluctuations, which favor the undistorted state. The total free energy is then
%ajmdec7 7 Embarassing algebra mistakes corrected!
\begin{equation}
F=-\left(\Delta-\Delta_0\right) \Theta\left(\Delta-\Delta_0\right)+\frac{1}{2}\frac{\Delta^2}{{\bar \Delta (T)}}
\label{Fdef}
\end{equation}
We see that for ${\bar \Delta}(T)<\Delta_0$, $F$ is minimized at $\Delta=0$ but for ${\bar \Delta}>\Delta_0$  the free energy is minimized at a value $\Delta>\Delta_0$ implying a Mott state, and that as ${\bar \Delta}$ continues to increase the magnitude of $\Delta$ and therefore the size of the Mott gap  further increases. Additional study of this issue using the realistic band structure and a better model of the phonon energetics would be desirable. 

Another material to which the considerations of this paper should apply is doped C$_{60}$. For this material, quantum chemical calculations suggest a $U \sim 1.5eV$, $J \approx 0.1eV $ \cite{Wierzbowska94,Gunnarsson97} and a bandwidth of about   $0.6eV$,\cite{Erwin03} corresponding in the language of this paper to $t\sim 0.15eV$ so $U \sim 10t$ and $J \sim U/15$.  Experimentally, A$_1$C$_{60}$ and A$_2$C$_{60}$ are insulators, while A$_3$C$_{60}$ is metallic and superconducting. The small $J$ limit of the theory is roughly consistent with the trend in behavior, with the interactions being strong enough to place the $n=1$ compound firmly within the Mott phase while the location of the $n=2,3$ materials is uncertain. The calculations presented here would suggest that A$_2$C$_{60}$ and A$_3$C$_{60}$ should be approximately equally likely to be insulating.  From this point of view, understanding the metallic nature of the A$_3$C$_{60}$ and insulating nature of A$_2$C$_{60}$ is an important open problem. 

\section{Summary}
%ajmdec7 added some text.t

In this paper we have used continuous-time quantum Monte Carlo methods to produce a comprehensive picture of the metal-insulator phase diagram and response to crystal fields of a ``three orbital" model which contains the essential physics of the fullerides, and the perovskite-based titanates, vanadates and ruthenates. We have documented the strong effect of the Hunds coupling on the location of the Mott transition and on the response to crystal fields and have placed a number of experimentally interesting materials on the phase diagram. The methods presented here provide a basis for detailed, material-specific calculations of realistic Hamiltonians. An important future direction for research is the investigation of the stability of the phases we have found against orbital and magnetic ordering. Work in this direction is in progress.

\acknowledgements

The calculations have been performed on the Hreidar and Brutus clusters at ETH Z\"urich, using the ALPS-library.\cite{ALPS} We acknowledge support from DOE ER-46169 (PW and AJM) and from the Swiss National Science Foundation (PW and EG).

\newpage
\clearpage
\newpage


\begin{thebibliography}{99}
\bibitem{Imada98} M. Imada, A. Fujimori and Y. Tokura,  Rev. Mod. Phys. {\bf 70}, 1039 (1998).
\bibitem{Science00} J. Orenstein and A. J. Millis, Science {\bf 288} 468, (2000).
\bibitem{Anderson86} P. W. Anderson, Science {\bf 235}, 1196 (1987).
\bibitem{Zhang88} F.C. Zhang, T.M. Rice, Phys. Rev., B {\bf 37}, 3759 (1988).
\bibitem{Gunnarsson96} O. Gunnarsson, E. Koch and R. M. Martin, Phys. Rev. {\bf B54} 11026 (1996).
\bibitem{Pavarini04} E. Pavarini, S. Biermann, A. Poteryaev, A. I. Lichtenstein, A. Georges, and O. K. Andersen, Phys. Rev. Lett. {\bf 92}, 176403 (2004).
\bibitem{Moore07} R. G. Moore, Jiandi Zhang, V. B. Nascimento, R. Jin, Jiandong Guo, G.T. Wang, Z. Fang, D. Mandrus, E. W. Plummer, Science {\bf 318}, 615 (2007).
\bibitem{Liebsch03} A. Liebsch, Phys. Rev. Lett. {\bf 91}, 226401 (2003).
\bibitem{Georges96} A. Georges, G. Kotliar, W. Krauth and M.~J. Rozenberg, Rev. Mod. Phys. {\bf 68}, 13 (1996).
\bibitem{Slaverotor} S. Florens and A. Georges, Phys. Rev. B 70, 035114 (2004); L. de' Medici, A. Georges, and S. Biermann, Phys. Rev. B {\bf 72} 205124 (2005); P. Lombardo, A.-M. Dar\'e, and R. Hayn, Phys. Rev. B {\bf 72} 245115 (2005).
\bibitem{Liebsch} C. A. Perroni, H. Ishida, and A. Liebsch, Phys. Rev. B {\bf 75}, 045125 (2007). 
\bibitem{deMedici08} L. de' Medici, S. R. Hassan, M. Capone, and X. Dai, arXiv:0808.1326.
%\bibitem{DeMedici08} Luca paper on multiorbital dmft--or should put this as a note added?
\bibitem{HirshFye1} J. E. Hirsch and R. M. Fye, Phys. Rev. Lett. {\bf 56}, 2521 (1986).
%\bibitem{Bluemer} N. Bl\"umer, PhD thesis, Universit¬at Augsburg (2002).
\bibitem{Rubtsov05} A. N. Rubtsov, V. V. Savkin and A. I. Lichtenstein, Phys. Rev. B {\bf 72}, 035122 (2005).
\bibitem{CTHirshFye} E. Gull, P. Werner, O. Parcollet, and M. Troyer, Europhys. Lett. {\bf 82}, 57003 (2008).
\bibitem{ED1} M. Caffarel and W. Krauth, Phys. Rev. Lett. {\bf 72}, 1545 (1994).
%\bibitem{ED2} Capone/DeMedici ED
\bibitem{NRG} R. Bulla, T. A. Costi, and T. Pruschke, Rev. Mod. Phys. {\bf 80}, 395 (2008).
%\bibitem{Sakai} S. Sakai, R. Arita, K. Held, and H. Aoki, Phys. Rev. B {\bf 74}, 155102 (2006).
\bibitem{Sakai} S. Sakai, R. Arita, and H. Aoki, Phys. Rev. B {\bf 70}, 172504 (2004).
\bibitem{Werner05} P.~Werner, A.~Comanac, L.~de' Medici, M.~Troyer and A.~J.~Millis, Phys. Rev. Lett. {\bf 97}, 076405 (2006).
\bibitem{Werner06} P.~Werner and A.~J.~Millis, Phys. Rev. B {\bf 74}, 155107 (2006). 
\bibitem{Werner07Crystal} P. Werner and A. J. Millis, Phys. Rev. Lett. {\bf 99}, 126405 (2007).
\bibitem{Haule07} K. Haule, Phys. Rev. B {\bf 75}, 155113 (2007).
\bibitem{ChrisChan08} C.-K. Chan {\it et al.}, in preparation.
\bibitem{Liebsch07chargetransfer} A. Liebsch, Phys. Rev. B {\bf 77}, 115115 (2008).
\bibitem{Fujioka08} J. Fujioka, S. Miyasaka, and Y. Tokura, Phys. Rev. B {\bf 77}, 144402 (2008).
\bibitem{Mikokawa96} T. Mizokawa and A. Fujimori, Phys. Rev. B {\bf 54}, 5368 (1996).
\bibitem{Sawada96} H. Sawada, N. Hamada, K. Terakura, and T. Asada, Phys. Rev. B {\bf 53}, 1272 (1996).
\bibitem{Werner08nfl} P. Werner, E. Gull, M.Troyer and A. J. Millis, Phys. Rev. Lett. {\bf 101},166405 (2008).
\bibitem{Cwik03} M. Cwik {\it et al.}, Phys. Rev. {\bf B68}, 060401 (2003)
\bibitem{Hemberger03}  J. Hemberger et. al., Phys. Rev. Lett., {\bf 91} 066403 (2003).
\bibitem{Okatov05} S. Okatov, A. Poteryaev and A. Lichtenstein, Europhys. Lett., {\bf 70} 499Ð505 (2005).
\bibitem{Mazin97} I. I. Mazin and D. J. Singh, Phys. Rev. Lett. {\bf 79}, 733 (1997).
\bibitem{Jung03} J. H. Jung, Z. Fang, J. P. He, Y. Kaneko, Y. Okimoto,  and Y. Tokura, Phys. Rev. Lett. {\bf 91}, 056403 (2003).
\bibitem{Liebsch07} A. Liebsch and H. Ishida, Phys. Rev. Lett. {\bf 98}, 216403 (2007).
\bibitem{Okamoto04} S. Okamoto and A. J. Millis, Phys. Rev. B {\bf 70}, 195120 (2004).
\bibitem{Wierzbowska94} M. Wierzbowska, M. Luders and E. Tosatti, J. Phys. B: At. Mol. Opt. Phys. {\bf 37}, 2685 (2004).
\bibitem{Gunnarsson97} O. Gunnarsson, Rev. Mod. Phys. {\bf 69} 575 (1996).
\bibitem{Erwin03} S. C. Erwin and W. E. Pickett, Science {\bf 254}, 842 (2003).



%\bibitem{Marianetti06} C.~A.~Marianetti, K.~Haule and O.~Parcollet, Phys. Rev. Lett. {\bf 99}, 246404 (2007).
%\bibitem{Liebsch07holepockets} A. Liebsch and H. Ishida, Eur. Phys. J. B {\bf 61}, 405 (2008).
%\bibitem{Werner07Doping} P.~Werner and A.~J.~Millis, Phys. Rev. B {\bf 75}, 085108 (2007).

%\bibitem{Kunes} J. Kunes, V. I. Anisimov, A. V. Lukoyanov, and D. Vollhardt, cond-mat/0612116.
%\bibitem{Georges05} F. Lechermann, S. Biermann and A. Georges, Phys. Rev. Lett. {\bf 94}, 166402 (2005).
%\bibitem{Nagaosascience} Y. Tokura and N. Nagaosa, Science {\bf 288}, 462 (2000).
%\bibitem{Liebsch03} A. Liebsch, Phys. Rev. Lett. , 226401 (2003).
%\bibitem{Koga04}  A. Liebsch, Phys. Rev. Lett. , 226401 (2003); A. Koga, N. Kawakami, T. M. Rice and M. Sigrist, Phys. Rev. Lett. 92, 216402 (2004).
%\bibitem{Okamoto}
%\bibitem{Sakai06} S. Sakai, R. Arita, K. Held, and H. Aoki, Phys. Rev. B {\bf 74}, 155102 (2006). 

%\bibitem{Laad01} M. S. Laad and E. M\"uller-Hartmann, Phys. Rev. Lett. {\bf 87}, 246402 (2001).
%\bibitem{Dodge07} J. S. Dodge {\it et al.}, arXiv:cond-mat/0006271.


%\bibitem{Rubtsov05} A. N. Rubtsov \textit{et al.}, V. V. Savkin and A. I. Lichtenstein, Phys. Rev. B {\bf 72}, 035122 (2005).
%\bibitem{Maier06} T. Maier, M. Jarrell, T. Pruschke, and H. M. Hettler, Rev. Mod. Phys. {\bf 78}, 865 (2006).
%\bibitem{Koga02} A. Koga, Y. Imai and N. Kawakami, Phys. Rev. B {\bf 66}, 165107 (2002).
%\bibitem{Fuhrmann06} A. Fuhrmann, D. Heilmann and H. Monien, Phys. Rev. B {\bf 73} 245118 (2006).
%\bibitem{Kancharla07} S. S. Kancharla and S. Okamoto, cond-mat/0703728.


\bibitem{ALPS} A. F. Albuquerque, F. Alet, P. Corboz, et al., Journal of 
Magnetism and Magnetic Materials {\bf 310}, 1187 (2007).
\end{thebibliography}
\end{document}